\documentclass{PoS}

\usepackage{amsmath}
\usepackage{amsfonts}
\usepackage{xspace}

\newcommand{\GeV}{{\,\,\rm GeV}}

\newcommand{\bea}{\begin{eqnarray}}
\newcommand{\eea}{\end{eqnarray}}
\newcommand{\bi}{\begin{itemize}}
\newcommand{\ei}{\end{itemize}}
\newcommand{\g}{\gamma}
\newcommand{\G}{\Gamma}

\newcommand{\ud}{\text{d}}

\newcommand{\babar}{\mbox{\ensuremath{{\displaystyle B}\!{\scriptstyle A}
{\displaystyle B}\!{\scriptstyle AR}}}\xspace}

\newcommand{\belle}{\mbox{\emph{Belle}}\xspace}

\title{Towards a global fit to extract the $B \to X_s \gamma$ decay rate and $\lvert V_{ub} \rvert$}
\ShortTitle{Towards a global fit to extract the $B \to X_s \gamma$ decay rate and $\lvert V_{ub} \rvert$}
\author{\speaker{Florian U. Bernlochner} \\
              Humboldt University of Berlin, 12489 Berlin, Germany\\
              E-Mail: \email{florian@slac.stanford.edu}}

\author{Heiko Lacker\\
              Humboldt University of Berlin, 12489 Berlin, Germany\\
              E-Mail: \email{lacker@physik.hu-berlin.de}}

\author{Zoltan Ligeti \\
              Ernest Orlando Lawrence Berkeley National Laboratory,
              University of California,\\ Berkeley, CA 94720, USA\\
              E-Mail: \email{ligeti@lbl.gov}}

\author{Iain W. Stewart \\
              Center for Theoretical Physics, Massachusetts Institute of Technology,\\Cambridge, MA 02139, USA\\
              E-Mail: \email{iains@mit.edu}}

\author{Frank J. Tackmann \\
              Center for Theoretical Physics, Massachusetts Institute of Technology,\\Cambridge, MA 02139, USA\\
              E-Mail: \email{frank@mit.edu}}

\author{Kerstin Tackmann \\
              CERN, CH-1211 Geneva 23, Switzerland\\
              E-Mail: \email{kerstin.tackmann@cern.ch}}

\author{(The \emph{SIMBA} Collaboration)}

\abstract{ The total $B \to X_s \gamma$ decay rate and the CKM-matrix element $\lvert V_{ub} \rvert$ play an important role in finding indirect evidence for new physics affecting the flavor sector of the Standard Model, complementary to direct searches at the LHC and Tevatron. Their determination from inclusive $B$-meson decays requires the precise knowledge of the parton distribution function of the $b$ quark in the $B$ meson, called the shape function. We implement a new model-independent framework for the shape function with reliable uncertainties based on an expansion in a suitable set of basis
functions. We present the current status of a global fit to \babar and \belle data to extract the shape function and the $B \to X_s \gamma$ decay rate. }

\FullConference{ HU-EP-10/61 \hspace{0.2cm} MIT--CTP 4195 \vspace{0.25cm} \\ 35th International Conference of High Energy Physics - ICHEP2010,\\July 22-28, 2010\\Paris France }

\begin{document}

\vspace{-0.5ex}
\section{Searching for new physics in the flavor sector}
\vspace{-1ex}

\begin{figure}[t!]
\vspace{-4ex}
\hfill\includegraphics[width=0.3\textwidth]{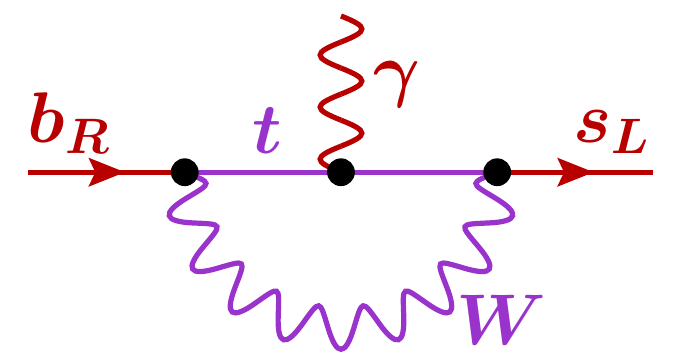}
\hfill\includegraphics[width=0.3\textwidth]{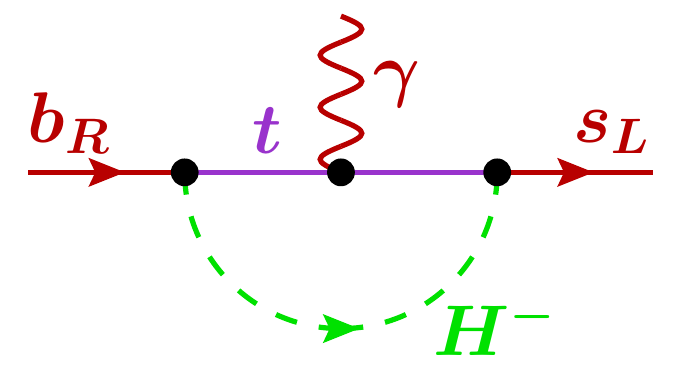}
\hspace*{\fill}
\caption{On the left-hand side a SM graph for the lowest order $B \to X_s\, \g$ decay is shown. In new-physics models with two Higgs doublets such as the minimal supersymmetric Standard Model the charged $W$-boson can be replaced by a charged Higgs, as shown on the right-hand side.}
\label{mssmext}
\vspace{-1ex}
\end{figure}

If physics beyond the Standard Model (SM) is present in the flavor sector, it can modify flavor changing neutral current processes  through the exchange of unknown virtual particles. The inclusive $B \to X_s \, \g$ decay via a radiative penguin diagram proves very sensitive to such corrections in many new-physics models, which could manifest itself in a modification of the total decay rate. The lowest order $B \to X_s \, \g$ decay and a new-physics contribution are shown in Fig. \ref{mssmext}. To constrain such contributions, a good knowledge of the SM prediction and precise measurements are needed. Hence, a model-independent framework, as proposed in Ref.~\cite{Ligeti:2008ac}, that combines all available information into a global fit is desirable to accurately determine the $B \to X_s \, \g$ decay rate. Here we present the current status and preliminary results of such a global fit to the available $B \to X_s \, \g$ data.

\vspace{-0.5ex}
\section{The $B \to X_s \, \g$ decay}
\vspace{-1ex}

The current analysis of the experimentally measured $B \to X_s \, \g$ branching fraction is performed by extrapolating the measurements
to the partial branching fraction for $E_\g > 1.6\GeV$, which yields~\cite{TheHeavyFlavorAveragingGroup:2010qj} $\mathcal{B}(E_{\g} > 1.6) =  (3.55 \pm 0.24 \pm 0.09) \times 10^{-4}$. This value is then compared to the fixed next-to-next-to-leading order (NNLO) SM prediction from Refs.~\cite{Misiak:2006ab, Misiak:2006zs}, $\mathcal{B}(E_{\g} > 1.6) =  (3.15 \pm 0.23) \times 10^{-4}$. The extrapolation still requires a theoretical calculation of the decay rate with the measured $E_\g$ cut. It assumes a model for the $b$-quark distribution function, the shape function, which introduces an unknown systematic uncertainty. In addition, the extrapolation only uses a single branching fraction from each experimental analysis, which is usually the one that has the smallest cut on $E_\g$ and correspondingly the largest experimental uncertainty. This means that only a small subset of all experimental information is used, and more precise measurements at higher values of $E_\g$ are not utilized.

A method that avoids these drawbacks was proposed in Ref.~\cite{Ligeti:2008ac}. Starting from the effective electroweak Hamiltonian, the $B \to X_s \, \g$ differential decay rate can be written as
\begin{align}
\label{eq:master}
\frac{ \ud \G_{B \to X_s \, \g} }{\ud E_\g}
&= \frac{G_F^2 \alpha_\mathrm{em}}{2\pi^4}\, E_\g^3\, (m_b^{1S})^2\,\lvert V_{tb} V_{ts}^* \rvert^2\,
\lvert C_7^\mathrm{incl}\rvert^2 \biggl[ \int\! \ud k\, \widehat W_{77}(k)\, \widehat F(m_B - 2E_\g - k)
\nonumber\\ & \quad
+ \sum_m W_m \, F_m(m_B - 2E_\g)   \biggr]
+ \sum_{i\neq 7} \mathcal{O}(C_7^\mathrm{incl} C_i) + \sum_{i,j\neq 7} \mathcal{O}(C_i C_j)
\,.\end{align}
The complete expressions entering Eq.~\eqref{eq:master} will be given in Ref.~\cite{Ligeti2010}. Here, $\widehat F(k)$ is the nonperturbative contribution to the $b$-quark distribution function in the $B$ meson. Its precise definition is given in Ref.~\cite{Ligeti:2008ac}. It is $\widehat F(k)$ that determines the shape of the $B \to X_s \, \g$ spectrum at large $E_\g$ and which we will refer to as the shape function. The function $\widehat W_{77}(k)$ contains the perturbative corrections to the spectrum resummed to next-to-next-to-leading-logarithmic order, and including the full NNLO corrections. At lowest order, $W_{77}(k) = \delta(k)$. The $F_m(k)$ are $1/m_b$ suppressed subleading shape functions. In a fit to $B \to X_s \, \g$ data only, they can be absorbed into $\widehat F(k)$ at lowest order in $\alpha_s$. The terms proportional to $C_{i\neq 7}$ are included at next-to-leading order for $i = 1,2,8$ using the SM values for $C_{1,2,8}$. They have almost no effect on the fit, because they are very small in the experimentally accessible region of the photon energy spectrum.

The coefficient $C_7^\mathrm{incl} = C_7^\mathrm{eff}(\mu_0) \overline m_b(\mu_0)/m_b^{1S} + \dotsb$, where $C_7^\mathrm{eff}(\mu_0)$ is the standard effective Wilson coefficient multiplying $O_7 = e/(16\pi^2)\, \overline{m}_b\, \bar s\, \sigma_{\mu\nu} F^{\mu\nu} P_R b$ in the electroweak Hamiltonian. The ellipses denote all virtual contributions from other operators that generate the same effective $b\to s\g$ vertex. This includes the terms that cancel the $\mu_0$ dependence of $C_7^\mathrm{eff}(\mu_0) \overline m_b(\mu_0)$, such that $C_7^\mathrm{incl}$ is $\mu_0$-independent to the order one is working at. In our analysis we consider $\lvert C_7^\mathrm{incl}\, V_{tb} V_{ts}^* \rvert$ as the unknown parameter that parametrizes the total $B\to X_s\,\g$ rate. It is extracted simultaneously with $\widehat F(k)$ from a fit to the measured photon energy spectra. This fitted value can then be compared to its SM prediction. In this approach all measurements contribute to constrain the $B\to X_s\,\g$ rate.

\vspace{-0.5ex}
\section{Extraction of the shape function $\widehat F(k)$ from data}
\vspace{-1ex}

\begin{figure}[t!]
\hfill\includegraphics[width=0.4\textwidth]{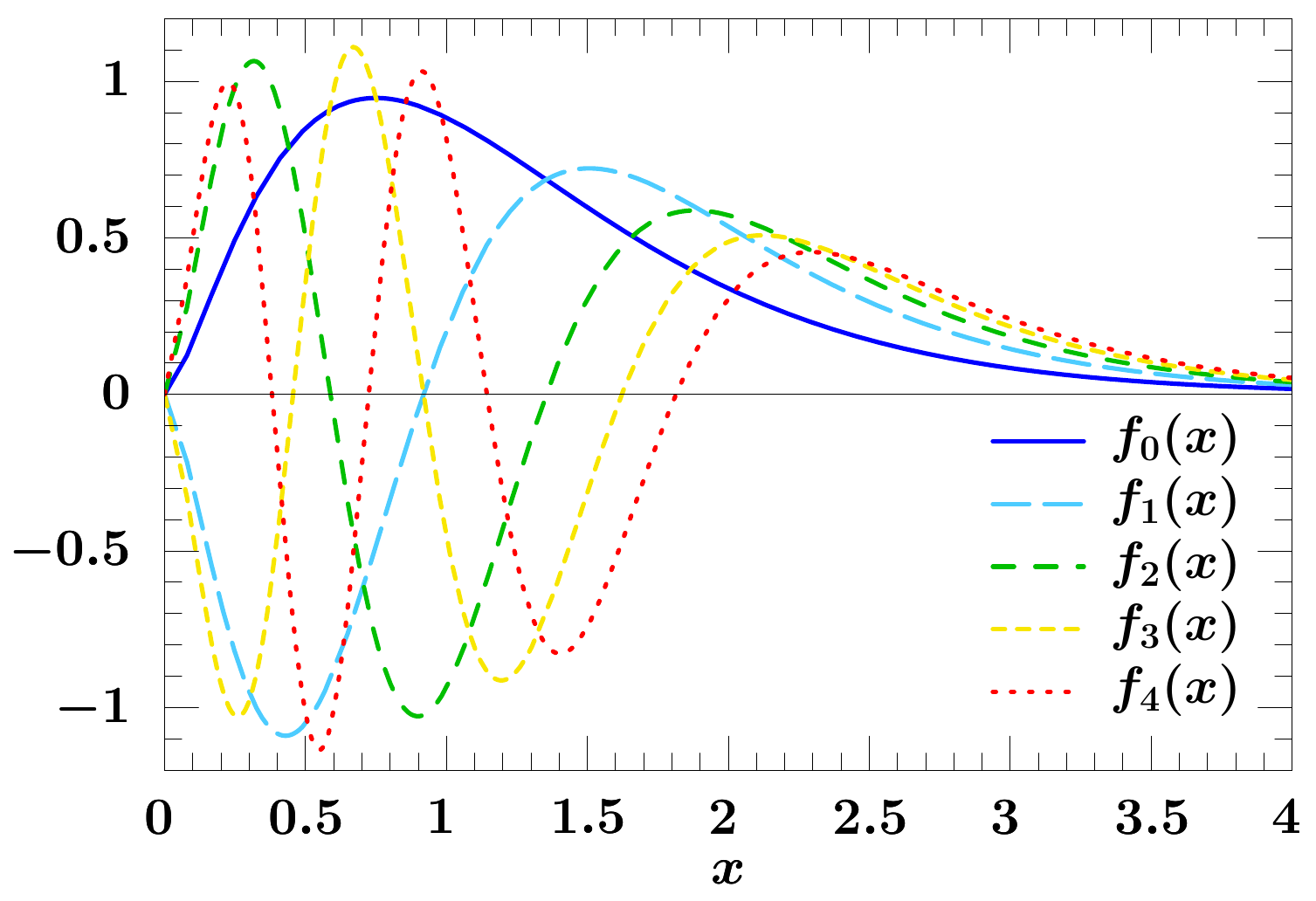}%
\hfill\includegraphics[width=0.41\textwidth]{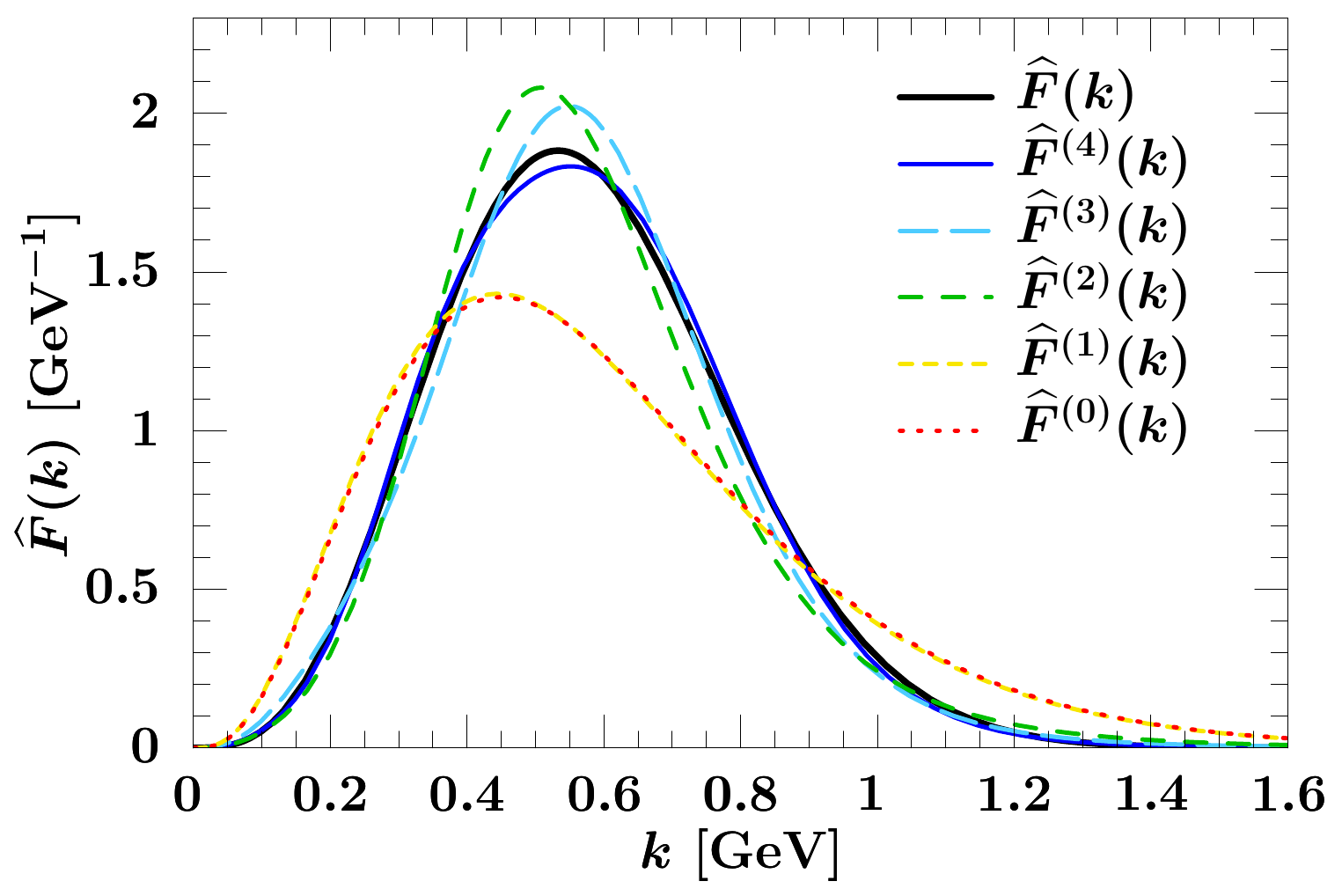}
\hspace*{\fill}
\vspace{-1ex}
\caption{Left: The first five basis functions $f_n(x)$. Right: Illustration of the convergence of the expansion. The black line is a Gaussian model function $\widehat F(k)$ and $\widehat F^{(N)}(k)$ are its expansions including up to $N+1$ terms.}   \label{funcbas}
\vspace{-1ex}
\end{figure}

The shape function can be expanded in a complete, orthogonal basis $f_n(x)$, constructed in Ref.~\cite{Ligeti:2008ac} and illustrated in Fig. \ref{funcbas} (where $\lambda \simeq \Lambda_\mathrm{QCD}$ is a parameter of the basis):
\begin{align}
  \widehat F(k) &= \frac{1}{\lambda} \biggl[ \sum_{n=0}^\infty \, c_n \,  f_n\Bigl(\frac{k}{\lambda}\Bigr) \biggr]^2
\qquad \text{with} \qquad
  \ \int \! \ud k \, \widehat F(k)  = \sum_{n=0}^{\infty} c_n^2 = 1 \, .
\end{align}
The shape of $\widehat F(k)$ is then parametrized by the basis coefficients $c_n$, which can be fitted from data. The experimental uncertainties and correlations in the measured spectra are captured in the uncertainties and correlations of the fitted coefficients $c_n$. In practice, the expansion must be truncated after $N$ terms, which introduces a residual model dependence from the chosen functional basis. The overall size of this truncation uncertainty scales as $1 - \sum_{n=0}^N c_n^2$ and the truncation order $N$ should be chosen such that this uncertainty is small in comparison to the experimental uncertainties of the fitted coefficients. In other words, we allow the available data to determine the precision to which the functional form of the shape function is known, by including as many basis coefficients in the fit as are required by the precision of the data. Hence, this approach allows for an experimental determination of the shape function with reliable and completely data-driven uncertainties.

\vspace{-0.5ex}
\section{Fit results}
\vspace{-1ex}

\begin{figure}[t!]
\hfill\includegraphics[width=0.33\textwidth]{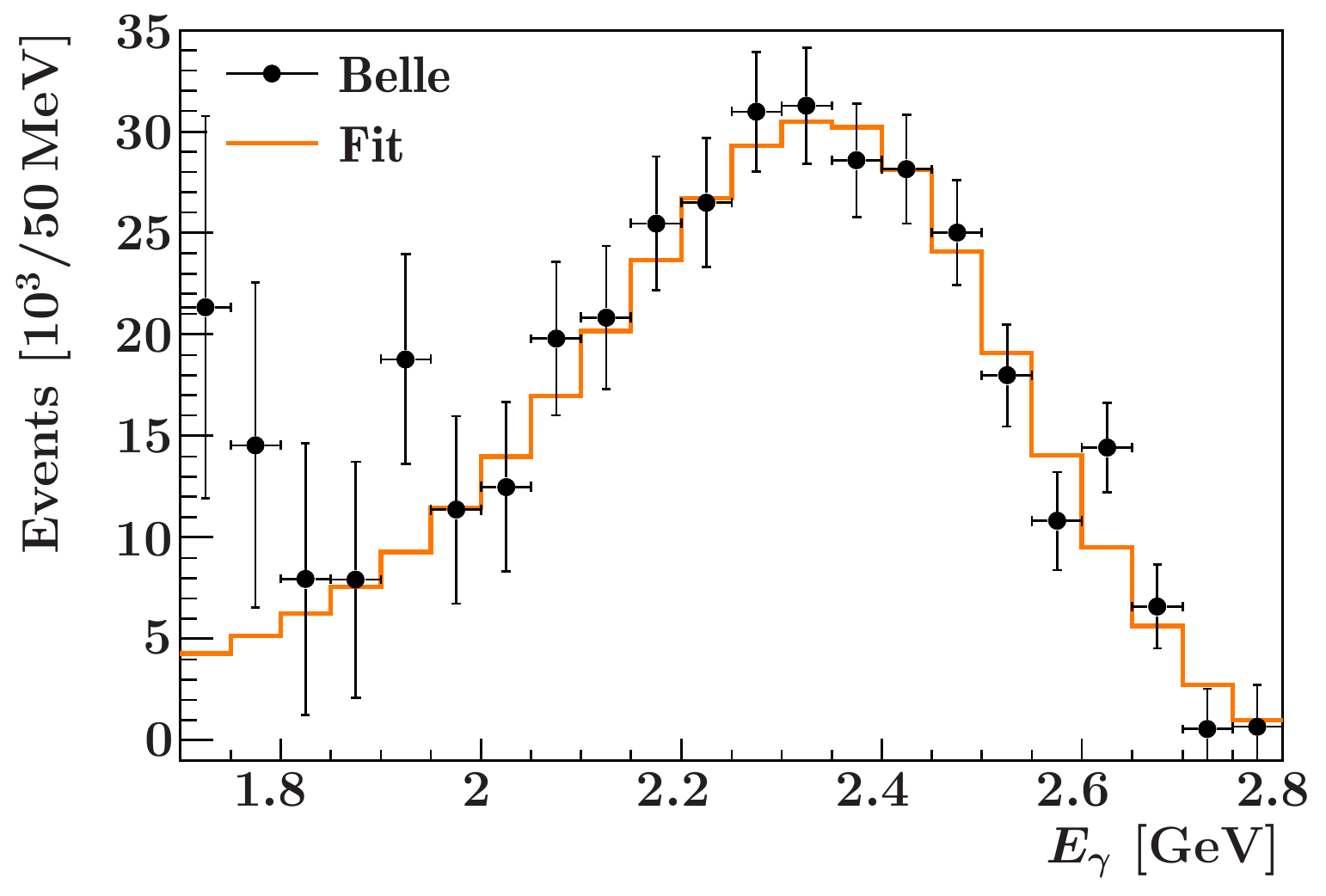}%
\hfill\includegraphics[width=0.33\textwidth]{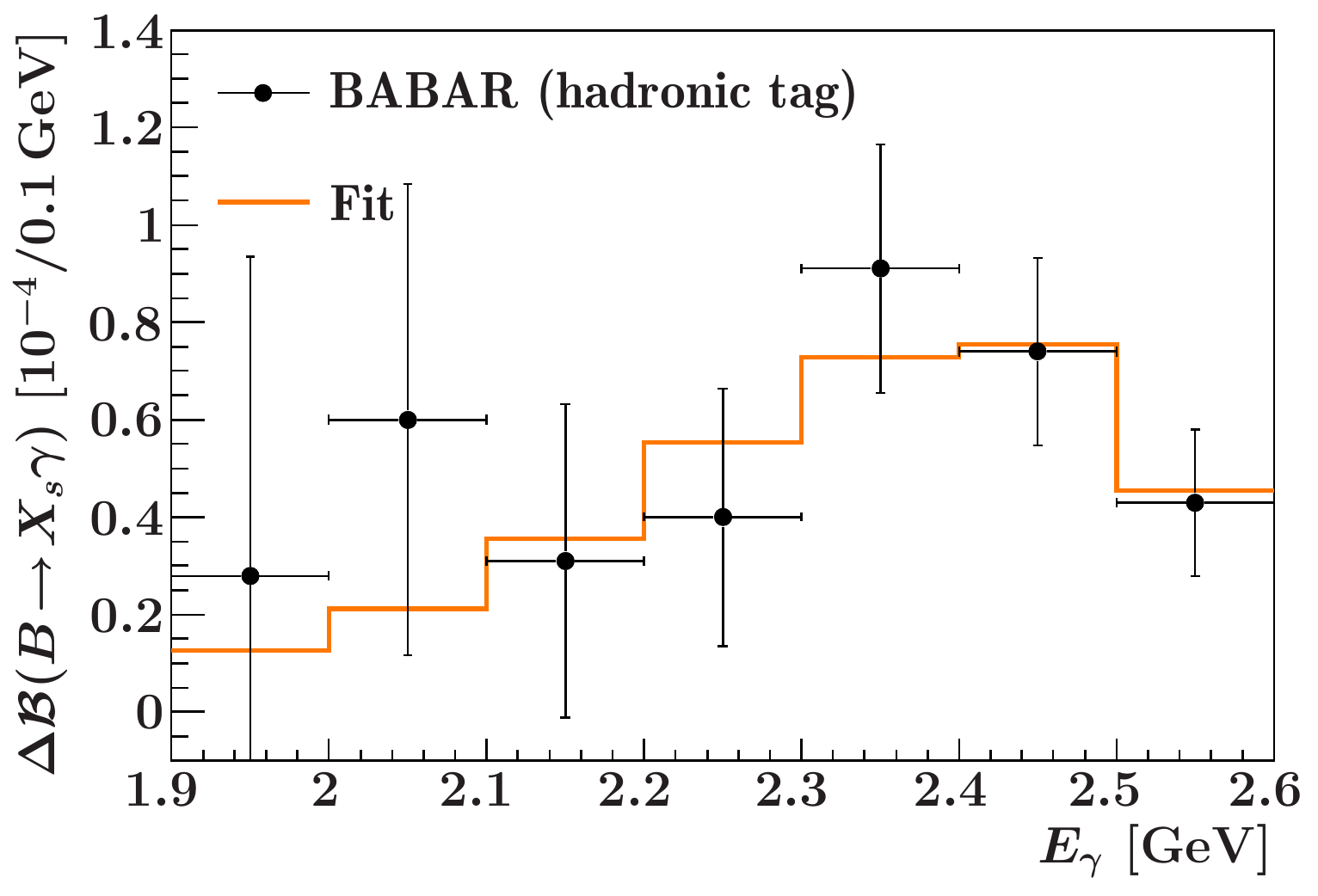}%
\hfill\includegraphics[width=0.33\textwidth]{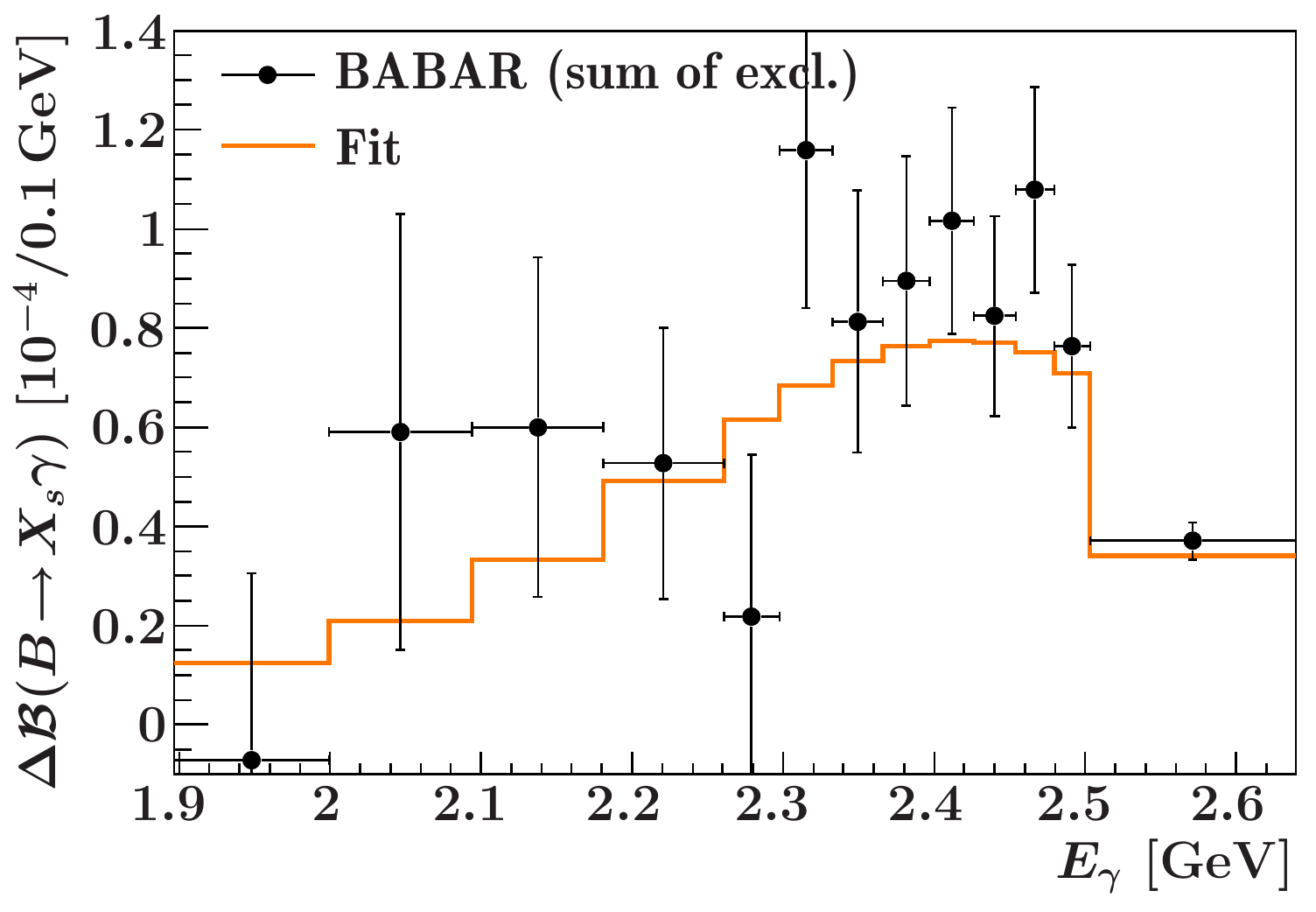}%
\hspace*{\fill}
\vspace{-1ex}
\caption{The \belle and \babar spectra from Refs.~\cite{:2009qg,Aubert:2007my,Aubert:2005cua}. The histograms show the result of the fit.} \label{fitresult}
\vspace{-1ex}
\end{figure}

\begin{figure}[t!]
\parbox{0.5\textwidth}{\includegraphics[width=0.5\textwidth]{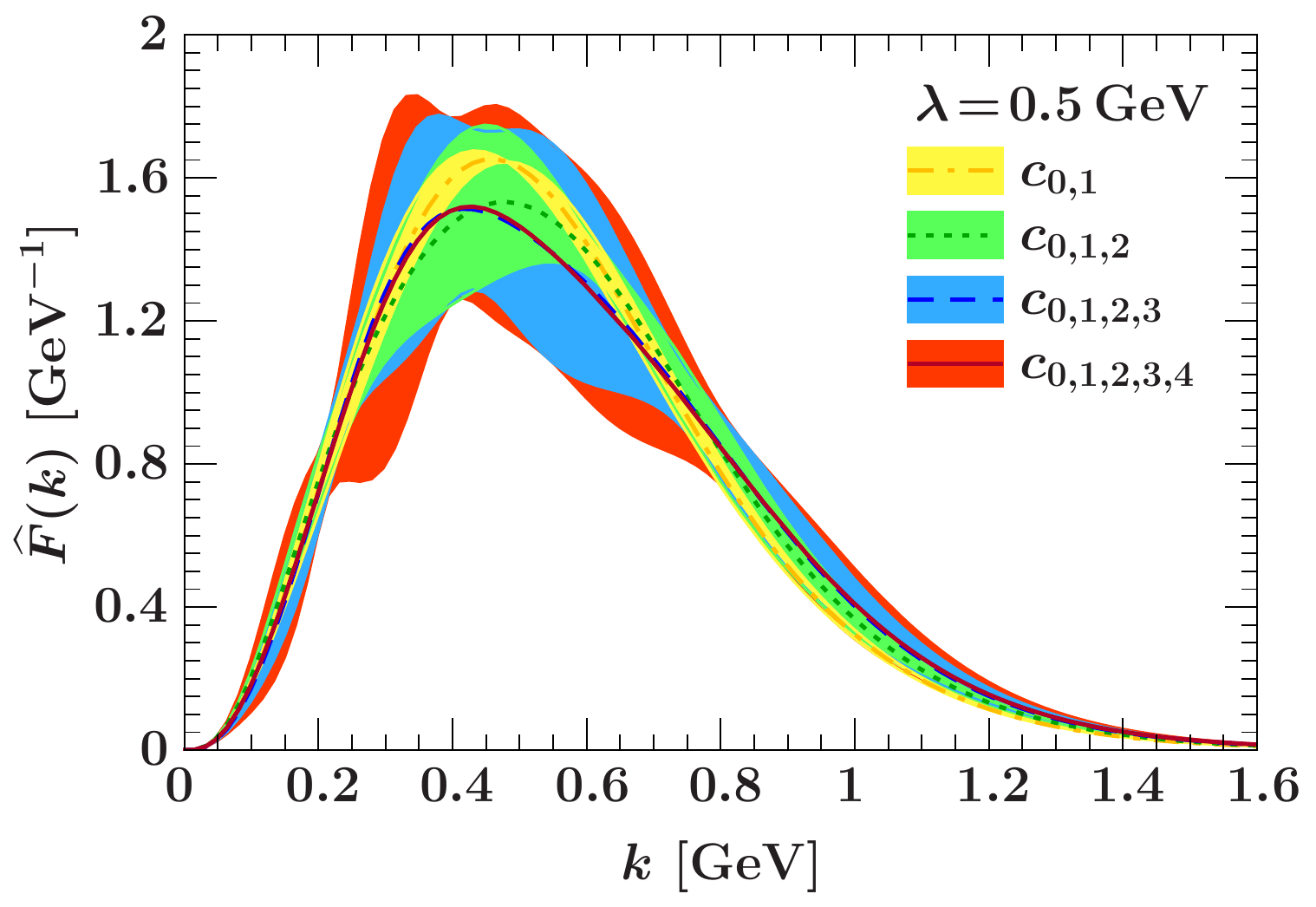}}%
\parbox{0.5\textwidth}{\includegraphics[width=0.5\textwidth]{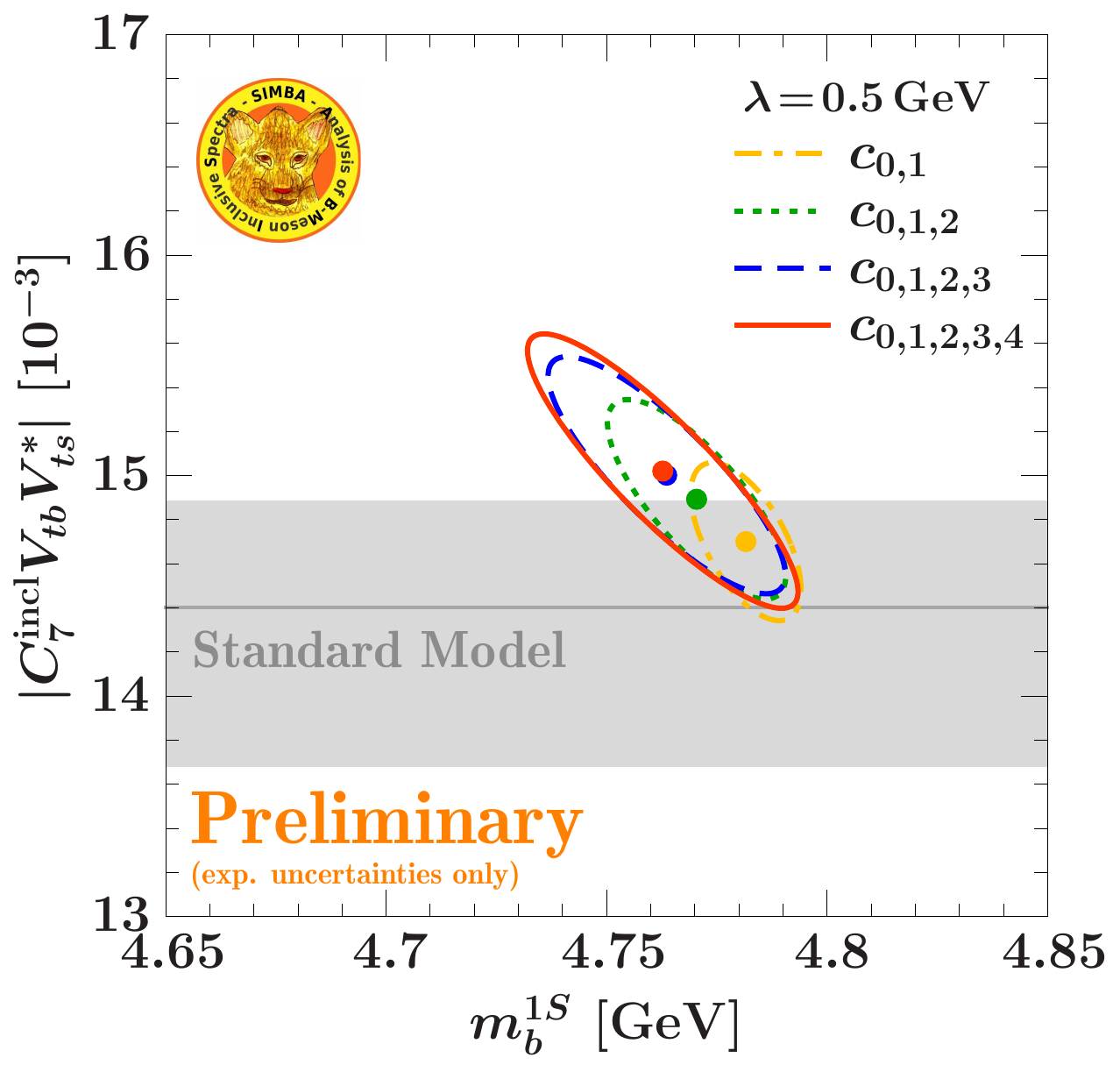}}
\vspace{-1ex}
\caption{Left: The extracted $\widehat{F}(k)$ (with absorbed $1/m_b$ corrections) for two ($c_{0,1}$), three ($c_{0,1,2}$), four ($c_{0,1,2,3}$), and five ($c_{0,1,2,3,4}$) coefficients and basis parameter $\lambda = 0.5\GeV$. The colored envelopes are given by the uncertainties and correlations of the extracted coefficients $c_n$. Right: The extracted values of $\lvert C_7^\mathrm{incl} \, V_{tb} V_{ts}^{*}\rvert$ and $m_b^{1S}$ with their respective $\Delta\chi^2 = 1$ contours. The grey band shows the SM value for $\lvert C_7^\mathrm{incl} \, V_{tb} V_{ts}^{*}\rvert$.}   \label{shapefunctionc7}
\vspace{-1ex}
\end{figure}

Using the above framework, we perform a global $\chi^2$ fit for $\widehat F(k)$ and $\lvert C_7^\mathrm{incl} V_{tb} V_{ts}^*\rvert$ to all available $B \to X_s \, \g$ photon energy spectra: The recent \belle measurement in Ref.~\cite{:2009qg}, and the two \babar measurements in Refs.~\cite{Aubert:2007my,Aubert:2005cua}. We have extensively tested our fitting procedure using pseudo-experiments. So far, the fit includes experimental uncertainties only. A preliminary study indicates that the theoretical uncertainties in the fit results are about the same size as the experimental ones.

The fit result using four basis coefficients and $\lambda = 0.5\GeV$ as the basis parameter is shown in Fig.~\ref{fitresult}. The fit has a $\chi^2 / \mathrm{dof} = 27.67/38$ and describes the measured spectra very well. The extracted shape function from fits with two to five coefficients are shown on the left in Fig.~\ref{shapefunctionc7}. On the right, we show the corresponding results for $\lvert C_7^\mathrm{incl}\, V_{tb} V_{ts}^{*}\rvert$ and $m_b^{1S}$, where the latter is computed from the moments of the extracted shape function. The fitted value for $\lvert C_7^\mathrm{incl}\, V_{tb} V_{ts}^{*}\rvert$ agrees within one standard deviation with its SM value, using $\lvert V_{tb} V_{ts}^{*}| = 40.68^{+0.4}_{-0.5} \times 10^{-3}$ and the next-to-leading order SM prediction for $C_7^\mathrm{incl} =  0.354^{+0.011}_{-0.012}$. For a more stringent comparison an evaluation of $C_7^\mathrm{incl}$ in the SM at NNLO using Refs.~\cite{Misiak:2006ab, Misiak:2006zs} would be very useful.

The results in Fig.~\ref{shapefunctionc7} show a convergent behavior as the number of basis functions is increased. The larger fit uncertainties with more coefficients originate from the larger number of degrees of freedom in the fit. A reliable value for the experimental uncertainties is given when the central values have converged and the last coefficient, here $c_4$, is compatible with zero within its uncertainties. At this point the truncation uncertainty can be neglected compared to the fit uncertainties in the coefficients. This also implies that using a fixed model function and fitting one or two model parameters will in general underestimate the true uncertainties of the shape function. We have also checked that using different basis parameters $\lambda = 0.4\GeV$ and $0.6\GeV$ yields consistent results.

\vspace{-0.5ex}
\section{Conclusions}
\vspace{-1ex}

We presented preliminary results from the first global fit for the nonperturbative shape function and the $B \to X_s \g$ decay rate, parametrized by $\lvert C_7^\mathrm{incl} \, V_{tb} \, V_{ts}^{*}\rvert$, within a model-independent framework. The extracted value of $\lvert C_7^\mathrm{incl} \, V_{tb} \, V_{ts}^{*}\rvert$ agrees with the SM prediction. From our fit we also determine $m_b^{1S}$. This constitutes the first step towards a global fit that combines all available data on both $B \to X_s \g$ and $B \to X_u\ell\bar \nu_l$ to determine $\lvert C_7^\mathrm{incl} \, V_{tb} \, V_{ts}^{*}\rvert$ and $\lvert V_{ub}\rvert$ in a model-independent way. 


\vspace{-1ex}
\section*{Acknowledgments}
\vspace{-1ex}

We are grateful to Antonio Limosani from \belle for providing us with the detector response matrix of Ref.~\cite{:2009qg}. We thank Francesca Di Lodovico from \babar, who provided us with the experimental correlations of Ref.~\cite{Aubert:2005cua}. This work was supported in part by the German Bundesministerium f\"ur Bildung und Forschung (F.B.), and the Director, Office of Science, Offices of High Energy and Nuclear Physics of the U.S.\ Department of Energy under the Contracts
DE-AC02-05CH11231 (Z.L.) and DE-FG02-94ER40818 (I.S. and F.T.).

\vspace{-1ex}
\bibliographystyle{jhep}
\bibliography{bibliography}
\vspace{-1ex}

\end{document}